# Generation of perfect-cavity-enhanced atom–photon entanglement with a millisecond lifetime via a spatially-multiplexed cavity

Minjie Wang, Shengzhi Wang, Tengfei Ma, Ya Li, Yan Xie, Haole Jiao, Hailong Liu, Shujing Li, Hai Wang[*]

The State Key Laboratory of Quantum Optics and Quantum Optics Devices, Institute of Opto-Electronics, Shanxi University, Taiyuan 030006
China Collaborative Innovation Center of Extreme Optics, Shanxi University, Taiyuan 030006, China

A qubit memory is the building block for quantum information. Cavity-enhanced spin-wave–photon entanglement has been achieved by applying dual-control modes. However, owing to cross readouts between the modes, the qubit retrieval efficiency is about one quarter lower than that for a single spin-wave mode at all storage times. Here, we overcome cross readouts using a multiplexed ring cavity. The cavity is embedded with a polarization interferometer, and we create a write-out photonic qubit entangled with a magnetic-field-insensitive spin-wave qubit by applying a single-mode write-laser beam to cold atoms. The spin-wave qubit is retrieved with a single-mode read-laser beam, and the quarter retrieval-efficiency loss is avoided at all storage times. Our experiment demonstrates 50% intrinsic retrieval efficiency for $540\,\mu s$ storage time, which is 13.5 times longer than the best reported result. Importantly, our multiplexed-cavity scheme paves one road to generate perfect-cavity-enhanced and large-scale multiplexed spin-wave–photon entanglement with a long lifetime.

**Introduction**

Atomic memory consisting of a qubit entangled with a photonic qubit forms the building block (repeater node) for long-distance quantum communications [1, 2] and large-scale quantum interworks [3-5] through a quantum repeater (QR) [1]. It may also be used for scalable linear-optical quantum computations (QC) [6]. In past decades, optical quantum memories (QMs) have been experimentally demonstrated with various matter systems [7] such as atomic ensembles [1, 8] and single-quantum systems including individual atoms [9,10], ions [11], and solid-state spins [12,13]. Compared with the QMs in a single-quantum system, the atomic-ensemble-based QMs feature collective enhancement and promise higher retrieval efficiencies [1]. For example, the highest retrieval efficiency for a single-photon qubit stored in a single atom trapped in a high-finesse optical cavity is 22% [10], while that in a high-optical-depth cold atomic ensemble is 85% [14]. With atomic ensembles, storage of quantum or classical light has been demonstrated via various schemes [8,15] including the Duan–Lukin–Cirac–Zoller (DLCZ) protocol [1,2, 16-37], electromagnetically-induced-transparency (EIT) dynamics [14, 38-46], photon echoes [47-54], Raman memory [55-58] and its variants [59, 60], etc. The DLCZ protocols create a spin-wave memory non-classically correlated [16-28] or entangled [29-37] with a Stokes photon via spontaneous Raman scattering by write pulses, which directly forms repeater nodes [1].

High retrieval efficiencies and long lifetimes are required for effectively achieving quantum information protocols [3, 7]. In long-distance entanglement distribution through QRs, a 1% increase in retrieval efficiency can improve the repeater rate by 10–14% [1]. In one-way QC, a qubit memory with retrieval efficiency exceeding 50% (66.7%) is required for error-correction [61] (photon-loss-tolerant [62]) protocols. However, the lifetime at 50% efficiency is defined as a benchmark of memory nodes in QRs [3, 25]. Additionally, establishing entanglement between two nodes separated by L=300 km in a "heralded" fashion requires storing a qubit for at least $L/c$=1.5 ms [7, 37, 63], with $c$ being the speed of light in fibers. To realize long-lived QMs, many studies have been done on decoherence of spin waves (SWs) in cold atomic ensembles [19-24, 31-34, 51]. These studies show that atomic-motion-induced decoherence can be suppressed either by lengthening SW wavelengths [20, 21, 34, 51] or confining the atoms in optical lattices [22-24, 32,33]. Inhomogeneous-broadening-induced decoherence may be suppressed using magnetic-field-insensitive (MFI) coherences to store SWs [20-24, 32-34]. Long-lived (0.1 s) atom–photon entanglement has been demonstrated by storing a memory qubit as two spatially-distinct MFI SWs in optical-lattice atoms [32]. However, the retrieval efficiency at zero delay only reached ~16% in that experiment. The retrieval efficiency for spin-wave qubits have been significantly improved using either high-optical-depth cold atoms [14, 45] or coupling moderate-optical-depth atoms with a low-finesse optical cavity [30]. Using the cavity scheme, Pan's group demonstrated intrinsic retrieval efficiency up to ~76% in entanglement generation between photonic polarization states and spin collective states. However, because the spin collective states include a magnetic-field-sensitive coherence, the memory lifetime is only ~30 μs [30]. Efficient and long-lived quantum correlation between a single-mode SW and a photon has been experimentally demonstrated by applying a single-control laser to optical-lattice atoms [24]. Because SW was stored as a MFI coherence and atom–light coupling enhanced by a ring cavity, that experiment demonstrated 50% efficiency at 50 ms and 76% efficiency at zero delay [24]. Furthermore, the same group applied dual-control laser beams to achieve cavity-enhanced entanglement between photonic and spin-wave qubits with sub second lifetimes [33]. However, because of the readout loss resulting from cross retrievals of the dual-control laser modes, the retrieval efficiency of the SW qubit was about one quarter lower than that of a single-mode SW at all storage times in that experiment [33]. For example, the efficiency of retrieving a qubit was ~58% at zero delay, about three quarters that of the single-mode one (~77%). In the present experiment, we overcome the readout loss of an SW qubit using a multiplexed ring cavity. The ring cavity is embedded with a polarization interferometer with cold atoms in its center. The two write-out fields, which individually propagate through the interferometer's arms, are created by applying a write laser beam to the cold atoms, and both resonate with the cavity. They form a photonic qubit entangled with MFI coherence and a long-wavelength spin-wave qubit. The spin-wave qubit is retrieved with a single-mode read laser, and the retrieval efficiencies for a qubit in our experiment are the same as those for a single-mode SW at any storage time. Our experimental results demonstrate intrinsic retrieval efficiencies of 50% and 66.7% for 540- and 230-μs storage times, respectively, which are 13.5 and 24 times higher than the best reported results [33] ([30]). The measured Bell parameter is $2.5 \pm 0.02$ at zero delay.

## RESULTS

Our experimental setup is shown in Fig. 1a. A cold atomic ensemble is centered in a polarization interferometer, which is formed by two beam displacers (BD1 and BD2). The interferometer is placed inside a ring cavity. Outside the interferometer (bottom of Fig. 1a), the cavity supports a single $TEM_{00}$ mode, labeled as cavity mode $A_{00}$. Inside the interferometer, the mode $A_{00}$ is spatially split into two cavity modes $A_{00}^L$ and $A_{00}^R$, which propagate along the left and right arms of the interferometer, respectively. We start the spin-wave-photon (atom-photon) entanglement generation after the atoms are prepared in the Zeeman state $|a, m_a = 0\rangle$ (Fig. 1b). At the beginning of a trial (see Materials and Methods for time sequence details), a $\sigma^+$-polarized write pulse red-detuned by 110 MHz to the $|a\rangle \to |e_1\rangle$ transition is applied to the atoms. This write pulse induces the Raman transition $|a, m_a = 0\rangle \to |b, m_b = 0\rangle$ via $|e_1, m' = 1\rangle$, which may emit $\sigma^+$-polarized Stokes photons and simultaneously create an SW excitation associated with the MFI coherence $|m_a = 0\rangle \leftrightarrow |m_b = 0\rangle$ (Fig. 1b). The Stokes (write-out) photon emitting into the interferometer arm $A_{00}^L$ ($A_{00}^R$) and moving along the cavity clockwise direction is denoted as photon $S_R$ ($S_L$). Accompanying the generation of photon $S_R$ ($S_L$), one collective excitation is created in the SW mode $M_R$ ($M_L$) defined by the wave-vector $k_{M_R} = k_W - k_{S_R}$ ($k_{M_L} = k_W - k_{S_L}$), where $k_W$ denotes the wave-vector of the write pulse and $k_{S_R}$ ($k_{S_L}$) that of the Stokes photon $S_R$ ($S_L$). The $\sigma^+$-polarized $S_R$ and $S_L$ fields are transformed into $H$-polarized fields by the quarter-wavelength ($\lambda/4$) plate QW$_S$. The $H$-polarized photon $S_L$ is transformed into a $V$-polarized photon by the half-wavelength ($\lambda/2$) plate HW$_S$. They are combined into the mode $A_{00}$ after BD2 and form a Stokes (write-out) qubit.

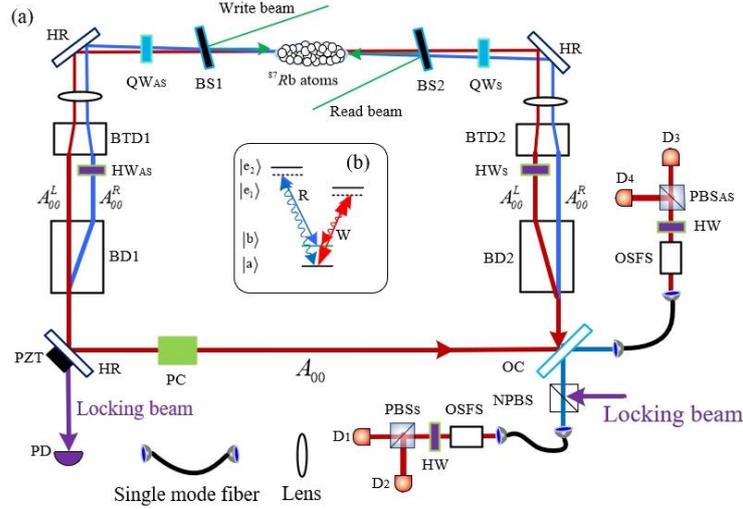

**Fig. 1.** Overview of the experimental setup (Fig. 1a) and relevant atomic levels (Fig. 1b). (a): A polarization interferometer formed by two beam displacers (BD1 and BD2) is inserted into a ring cavity. The cavity is formed by three mirrors (HR) that are highly reflective at 795 nm and an output coupler (OC) with a refection rate of 80%. The cold $^{87}$Rb atoms are centered at the interferometer. Inside the polarization interferometer, the two optical modes are the cavity modes $A_{00}^L$ and $A_{00}^R$ propagating in the interferometer's arms and marked by red and blue lines, respectively. Outside the interferometer, the two optical fields, one of which is H-polarized and the other V-polarized, are combined into a spatial mode A$_{00}$. Using the phase compensator (PC), we eliminate phase difference between $A_{00}^L$ and $A_{00}^R$ modes and then make them resonance with the cavity [34]. The length of the ring cavity is 6000 mm, corresponding to a free spectral range of 50 MHz. The cavity has finesses of 16.9 and 17.0 for the light-field modes $A_{00}^L$ and $A_{00}^R$. A 795-nm laser beam is coupled to the cavity modes with a nonpolarized beam splitter (NPBS) to lock the cavity length. A piezoelectric transducer (PZT) is used to lock the cavity by displacing an HR mirror via a PID control loop. PBS: Polarizing beam splitter; PD: photoelectric detector; BTD: beam transformation device [34] with a transformation factor of 2; OSFS: optical-spectrum-filter set [34]; BS1 (BS2): non-polarizing beam splitter with a reflectance of 1% (3%). (b): Relevant atomic levels.

The modes $M_R$ and $M_L$ form a spin-wave qubit, which is entangled with the Stokes qubit [34]. The atom–photon joint state is written as

$$\rho_{ap} = |0\rangle\langle 0| + \sqrt{\chi}|\Phi_{a\text{-}p}\rangle\langle\Phi_{a\text{-}p}|, \quad (1)$$

where $|0\rangle$ is the vacuum component, $\chi$ ($\ll 1$) represents the probability of creating the entangled atom–photon pair in each trial, $\Phi_{a\text{-}p} = |H\rangle_S|M_R\rangle + e^{i\varphi}|V\rangle_S|M_L\rangle$ denotes the spin-wave-photon entanglement state, $|H\rangle_S$ ($|V\rangle_S$) the H- (V-) polarized Stokes photon, $|M_R\rangle$ ($|M_L\rangle$) the SW excitation in the mode $M_R$ ($M_L$), and $\varphi$ the phase difference between the $S_R$ and $S_L$ fields. After a storage time $t$, we apply a $\sigma^-$-polarized read pulse that is red-detuned by 110 MHz to the $|b\rangle \rightarrow |e_2\rangle$ transition and counter-propagates with the write beam to convert the spin wave $|M_R\rangle$ ($|M_L\rangle$) into an anti-Stokes photon $AS_R$ ($AS_L$). The retrieved photon $AS_R$ ($AS_L$) is $\sigma^-$-polarized and emitted into the spatial mode determined by the wave-vector constraint $k_{AS_R} \approx -k_{S_R}$ ($k_{AS_L} \approx -k_{S_L}$); i.e., it propagates in the arm $A_R$ ($A_L$) in the opposite direction to the $S_R$ ($S_L$) photon. The $\sigma^-$-polarized fields $AS_R$ and $AS_L$ are transformed into $H$-polarized fields by the λ/4 plate QW$_{AS}$. The $H$-polarized field $AS_L$ is transformed into a $V$-polarized field by the λ/2 plate HW$_{AS}$. They are combined into mode $A_{00}$ after BD1 and form an anti-Stokes qubit. Thus, the atom–photon state $\Phi_{a\text{-}p}$ is transformed into the two-photon entangled state $\Phi_{pp}=|H\rangle_S|H\rangle_{AS}+e^{i(\varphi+\psi)}|V\rangle_S|V\rangle_{AS}$, where $\psi$ is the phase difference between the anti-Stokes fields $AS_R$ and $AS_L$ before they overlap at BD1. Using a phase compensator (PC in Fig. 1a), we set the phase difference $\varphi+\psi$ to zero.

By precisely tuning the frequency detuning of the write and read laser beams, we make the Stokes and anti-Stokes fields both resonate with the ring cavity. This enhances the atom–photon interactions, which allows an increase in the retrieval efficiency through the Purcell effect. The transmission of the write-out (readout) photon from the cavity output coupler passes through a sing-mode fiber and then is guided into a polarization-beam splitter labeled as $\text{PBS}_S$ ($\text{PBS}_{AS}$) in Fig. 1a. The two outputs of $\text{PBS}_S$ ($\text{PBS}_{AS}$) are sent to single-photon detectors $D_1$ ($D_3$) and $D_2$ ($D_4$). The polarization angle $\theta_S$ ($\theta_{AS}$) of the Stokes (anti-Stokes) field is changed by rotating a λ/2 plate before $\text{PBS}_S$ ($\text{PBS}_{AS}$).

The intrinsic retrieval efficiency of the SW qubit can be measured as $R_{qu}^{inc}=P_{S,AS}/(\eta_{TD}P_S)$, where $P_{S,AS}=P_{D_1,D_3}+P_{D_2,D_4}$; $P_{D_1,D_3}$ ($P_{D_2,D_4}$) is the probability of detecting a coincidence between the detectors $D_1$ ($D_2$) and $D_3$ ($D_4$) for $\theta_S=\theta_{AS}=0°$, which refers to the H- (V-) polarized coincidences between the Stokes and anti-Stokes photons; $P_S=P_{D_1}+P_{D_2}$, where $P_{D_1}$ ($P_{D_2}$) is the probability of detecting a Stokes photon at $D_1$ ($D_2$); $\eta_{TD}=\eta_{cav}\eta_t\eta_D$ is the total detection efficiency of the read-out (anti-Stokes) channel, which includes the efficiency of light escaping from the ring cavity, $\eta_{esp} \approx 60\%$, the transmission efficiency from the cavity to the detectors, $\eta_t \approx 36\%$ (see Materials and Methods for details), and the detection efficiency of the single-photon detectors, $\eta_D \approx 68\%$. Thus, the total detection efficiency is $\eta_{TD} \approx 15\%$. Moreover, the intrinsic retrieval efficiency for an individual SW $M_L$ ($M_R$) mode is defined as $R_L^{inc} = P_{D_1,D_3}/(\eta_{TD} P_{D_1})$ ($R_R^{inc} = P_{D_2,D_4}/(\eta_{TD} P_{D_2})$). Fig. 2 plots the measured efficiencies $R_{qu}^{inc}$ (red circle dots), $R_L^{inc}$ (blue square dots), and $R_R^{inc}$ (green triangle dots) as functions of storage time $t$. From the figure, we see that $R_{qu}^{inc} \approx R_L^{inc} \approx R_R^{inc}$ for different times $t$, which means that the retrieval efficiency for an SW qubit is the same as that for a single-mode SW. This shows that the efficiency loss of retrieving the qubit, which is a key limit in state-of-the-art work [33], is overcome in our experiment. The solid grey curve is the fit to the retrieval efficiencies $R_{qu}^{inc}$, $R_L^{inc}$, and $R_R^{inc}$ according to the function $R^{inc}(t) = R_0 \left( e^{-t^2/\tau_0^2} + e^{-t/\tau_0} \right)/2$, which yields a zero-delay retrieval efficiency $R_0 = 77\%$ and $\tau_0 \approx 1ms$, resulting in $R_{inc}(t = 0.23\text{ms}) \approx 66.7\%$ and $R_{inc}(t = 0.54\text{ms}) \approx 50\%$.

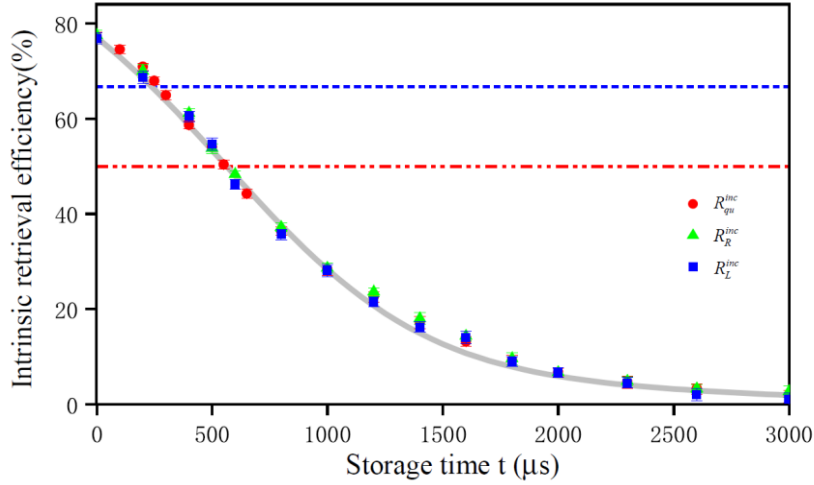

**Fig.2** Intrinsic retrieval efficiencies as a function of storage time $t$ for $\chi =1\%$. Error bars represents 1 standard deviation.

Next, we measure the Clauser–Horne–Shimony–Holt (CHSH) inequality, which is a type of Bell inequality, to confirm the spin-wave-photon entanglement state $\Phi_{\text{a-p}}$. The Bell CHSH parameter is defined as $S_{Bell} = |E(\theta_S, \theta_{AS}) - E(\theta_S, \theta'_{AS}) + E(\theta'_S, \theta_{AS}) + E(\theta'_S, \theta'_{AS})| < 2$ with the correlation function $E(\theta_S, \theta_{AS})$, which is written as

$\frac{C_{13}(\theta_S,\theta_{AS}) + C_{24}(\theta_S,\theta_{AS}) - C_{14}(\theta_S,\theta_{AS}) - C_{23}(\theta_S,\theta_{AS})}{C_{13}(\theta_S,\theta_{AS}) + C_{24}(\theta_S,\theta_{AS}) + C_{14}(\theta_S,\theta_{AS}) + C_{23}(\theta_S,\theta_{AS})}$. For example, $C_{13}(\theta_S,\theta_{AS})$ ($C_{24}(\theta_S,\theta_{AS})$) denotes the coincidence counts between the detectors $D_1$ ($D_2$) and $D_3$ ($D_4$) for the polarization angles $\theta_S$ and $\theta_{AS}$. We used the canonical settings $\theta_S = 0°$, $\theta'_S = 45°$, $\theta_{AS} = 22.5°$, and $\theta'_{AS} = 67.5°$ in measuring the Bell parameter $S_{Bell}$. Fig. 3 shows the decay of $S_{Bell}$ as a function of storage time $t$ (blue squares) for $\chi = 2\%$. At $t \approx 0\,\mu s$, $S_{Bell} = 2.5 \pm 0.02$, while at $t = 1.15$ ms, $S_{Bell} = 2.05 \pm 0.03$. These violate the Bell inequality by 25 and 1.7 standard deviations, respectively. Furthermore, at $t=2.6$ ms, $S_{Bell} = 1.15 \pm 0.03$, which corresponds to a fidelity of $F \approx 0.55$ and exceeds the bound of 0.5 required to observe entanglement for a Bell state.

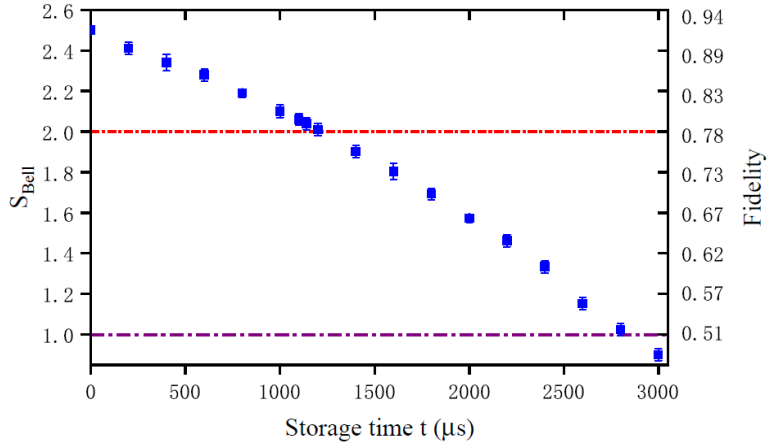

**Fig. 3.** Measured Bell parameter as a function of storage time *t* for $\chi = 2\%$. Error bars represent 1 standard deviation.

**DISCUSSION**

To generate the long-lived qubit memory entangled with a photonic qubit via the DLCZ protocol, a key step is to store the memory qubit as two MFI spin-wave modes [32-34]. The scheme used in Ref. [33] is not suitable for cavity-enhanced retrieval because it requires duplicating the cavity setup, which makes the whole system overcomplicated. The experiment in Ref. [33] creates two MFI spin-wave modes by applying dual write-laser beams to optical-lattice atoms, which in turn requires applying dual read-laser beams to retrieve the spin waves. The cross readouts induced by the two read-laser beams significantly affect the cavity-enhanced readout in that experiment [33]. Using the long-lived qubit memory in Ref. [34], our experiment consists of a ring cavity embedded with a polarization interferometer and demonstrates cavity-enhanced spin-wave-photon entanglement with a memory qubit stored as an MFI coherence. Two spatial modes of the write-out (readout) photons are created (retrieved) by a single-control laser model, and all resonate with the ring cavity. This ensures the cross readouts are avoided, resulting in cavity- perfect-enhanced (CPE) retrieval.

Compared with previous works on cavity enhanced retrieval [21,24,30, 33,58], our experiment has the advantage that the cavity simultaneously resonates with two spatially distinct photon modes, which makes it possible to generate perfect-cavity-enhanced and large-scale multiplexed spin-wave–photon entanglement with a long lifetime (see Materials and Methods for details).

**Comparison with other spin-wave qubit memories.**

Fig. 4 plots the measured retrieval efficiencies as functions of the storage time in various qubit memory systems via different storage schemes. The shaded region shows the results for an ideal-optical-fiber loop. Curves (a) and (b) are the results for storages of single-photon [14] and weak-coherent-light [45] qubits via an EIT in high-optical-density cold atoms, with the efficiencies reaching 85% and 68% at zero delay, respectively. However, the memory lifetimes all are a dozen seconds. Curve (c) represents the results for qubit memory stored as MFI spin waves via EIT in cold atoms [40], which feature a millisecond lifetime but a relatively low efficiency (8%) at zero delay. Curve (d) represents single-photon qubit storage via EIT in single atoms [10], which has a storage efficiency of 22% and lifetime of 100 ms. Curves (e)–(h) represent qubit memory based on the DLCZ protocol [2]. Curve (e) represents the results for a highly reversible [30] but short-lived atom–photon entanglement source, while (f) is for a long-lived but weakly reversible source [32]. Curves (g) and (h) are the results of Ref. [33] and our experiment. Our experimental result of 50% (66.7%) intrinsic retrieval efficiency for a storage time of 540 μs (230 μs) is 13.5 (24) times higher than the previously reported best result, which is in Ref. [33] (Ref. [30]).

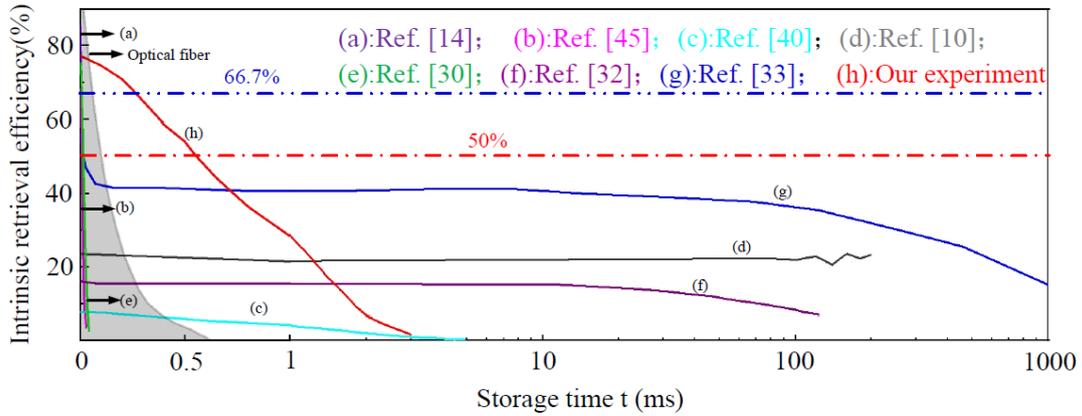

**Fig. 4.** Measured retrieval efficiencies as functions of storage time t in various memory systems.

**Possible improvements of lifetime**.

As for the limit to atomic-motion-induced dephasing in cold atoms, the $1/e$ lifetime in our experiment is 1 ms, which is far less than the ~460-ms lifetime in Ref [33]. However, this is not a fundamental limit. By loading the cold atoms into optical lattices and using magic field values, our experiment can extend the 1/*e* lifetime to that value. Furthermore, in an optical EIT storage [64], a 1/*e* lifetime of 6.9 s or 16 s has been achieved by further minimizing magnetic field gradients or using dynamic decoupling pulse sequences; therefore, quantum memories with comparable storage times may be within reach.

**Prospective applications of atom-photon entanglement QI in QR.**

Entanglement generations between two remote nodes (quantum memories) belonging to individual repeater links have been demonstrated with cold atomic ensembles [26, 31, 65-67], crystals [68, 69], and single-quantum systems such as atoms [70, 71], quantum dots [12, 13], etc. The next important challenges for developing ground-based QRs include deterministically generating entanglement in individual repeater links over a significant spatial distance, connecting two or more repeater links via entanglement swapping, and achieving a quantum repeater architecture with a higher entanglement distribution rate than direct transmission [3]. A previous experiment has demonstrated the longest spatially separated elementary link by entangling two cold-atomic-ensemble memories over 22 km of field-deployed fibers [31] via two-photon interference [72]. However, entanglement in that experiment was probabilistically generated. The required time for deterministically generating entanglement in such a link is evaluated as ~150 s [31], which is far beyond the state-of-the-art $1/e$ lifetime of ~16 s reported in optical EIT storage in optical-lattice atoms [64]. Thus, it remains challenging to make QRs using cold atoms as nodes. Large-scale multimode quantum storages have been proposed to overcome this challenge [3, 63, 73]. The DLCZ protocol in cold atoms has been used to demonstrate multiplexed storages of qubits with hundreds of spatial modes [36, 74] and tens of temporal modes [75], which allow us to achieve multiplexed qubit memories with thousands of modes by combining both schemes [52, 76]. Fig. 5 shows the repeater rate we calculated for distributing an entangled photon pair over distance $L$ through a four-level QR via two-photon interference for the following cases: high-performance quantum technologies including large-scale-multiplexing (e.g., 1000 modes), cavity-enhanced-retrieval and long-lifetime (16 s) [64] quantum storages, and high-efficiency memory-to-telecom frequency conversion (with an efficiency of 33%) [31] integrated into individual atom–photon entanglement sources used as nodes. The blue solid and red dashed curves are the results for CPE retrieval ($R_0$=77%) and cavity-imperfectly-enhanced (CIE) retrieval ($R_0$=58%), respectively. For a fixed repeater rate of $10^{-4}$, one can see that the QR using CPE-retrieval nodes may achieve an entangled photon pair distribution over $L \approx 1000 km$, while that using CIE retrieval nodes may only achieve entanglement distribution over $L \approx 430 km$, showing a significant advance of CPE over CIE retrieval schemes. In conclusion, our presented experiment demonstrates a perfectly cavity-enhanced and long-lived spin-wave–photon entanglement, thus benefitting quantum networks for long-distance quantum communications and large-scale quantum computations.

*Corresponding author: wanghai@sxu.edu.cn

**Acknowledgment**

We acknowledge funding support from Key Project of the Ministry of Science and Technology of China (Grant No. 2016YFA0301402); The National Natural Science Foundation of China (Grants: No. 11475109, No. 11974228), Fund for Shanxi "1331 Project" Key Subjects Construction.

**Supplementary Materials**

**Experimental details.** The experiment is performed in a cyclic fashion. In each experimental cycle, the durations for preparing cold atoms and running the experiment for spin-wave-photon entanglement (SWPE) generation are 42 ms and 8 ms, respectively, corresponding to a 20-Hz cycle frequency. During the preparation stage, more than $10^8$ atoms of $^{87}$Rb are trapped in a two-dimensional magneto-optical trap (MOT) for 41.5 ms and further cooled via Sisyphus cooling for 0.5 ms. The cloud of cold atoms has a size of ~5×2×2 mm$^3$, a temperature of ~100 μK, and an optical density of ~16. At the end of this preparation stage, a bias magnetic field of $B_0$=4 G is applied along the z-axis (see Fig. 1a), and the atoms are optically pumped into the initial level $|5^2S_{1/2}, F=1, m=0\rangle$. After the preparation stage, the 8-ms experimental run containing a large number of SWPE-generation trials starts. At the beginning of a trial, a write pulse with a duration of ~300 ns is applied to the atomic ensemble to generate correlated pairs of Stokes photons and spin-wave excitations. The detection events at the Stokes detectors $D_1$ and $D_2$ in Fig. 1a are analyzed with a field-programmable gate array (FPGA). As soon as a Stokes photon qubit is detected by either one of these detectors, SWPE is generated and the FPGA sends out a feed-forward signal to stop the write processes. After a storage time $t$, a read laser pulse with a duration of ~300 ns is applied to the atoms to convert the spin-wave qubit into the anti-Stokes photon qubit.

After a 1300-ns interval, a cleaning pulse with a duration of 200 ns is applied to pump the atoms into the initial level $\left|5^2 S_{1/2}, F=1, m=0\right\rangle$. Then, the next SWPE-generation trial starts. However, in most cases, the Stokes photon is not detected during the write pulse owing to the low excitation probability ($\chi \leq 2\%$). If this is the case, the atoms are pumped directly back into the initial level by the read and cleaning pulses. Subsequently, the next trial starts, i.e., the write pulse is applied. The delay between the two adjacent write pulses for a storage time of $t \approx 1$ μs is 2000 ns. Therefore, the 8-ms experimental run contains ~4000 experimental trials. Considering that a 1-s experiment contains 20 cycles, the repetition rate of the SWPE-generation trail is $r = 8 \times 10^4$.

At the center of the atoms, the waist diameter of the $A_{00}^L$ and $A_{00}^R$ cavity modes are both ~0.5 mm. The powers of the two beams are ~300 μW and ~10 mW, respectively. The read laser is red-detuned by 110 MHz to the transition $|b\rangle \rightarrow |e_2\rangle$. To block the write (read) laser beam in the Stokes and anti-Stokes channels, we place an optical-spectrum-filter set [35] (OSFS) before each polarization beam splitter PBS$_S$ (PBS$_{AS}$). Each OSFS comprises five Fabry–Perot etalons, which attenuate the write (read) beam by a factor of ~$8.1 \times 10^{-12}$ (~$3.8 \times 10^{-11}$) and transmit the Stokes (anti-Stokes) fields with a transmission of ~56%. Additionally, in the Stokes (anti-Stokes) detection channel, the spatial separation of the Stokes (anti-Stokes fields) from the strong write (read) beam attenuates the write (read) beam by ~$10^{-4}$. In this experiment, we measured the uncorrelated noise probability in the anti-Stokes mode, which is $p_N \approx 10^{-4}$ per read pulse (300 ns). Such uncorrelated noise mainly results from the leakage of the read beam into the anti-Stokes detection channel.

The efficiency of light escaping from the ring cavity is defined as $\eta_{esp} = \frac{T_{OCM}}{T_{OCM} + L}$, where $T_{OCM} = 20\%$ is the transmission of the output coupler mirror, and $L \approx 13\%$ is the cavity loss. Thus, we have $\eta_{esp} \approx 60\%$.

In the readout channel, the transmission efficiency $\eta_t$ of the photons from the cavity to the detectors includes the coupling efficiency $\eta_{SMF} \approx 0.71$ of the single-mode fiber SMF$_{AS}$, the transmission $\eta_{Filter} \approx 0.56$ of the optical-spectrum-filter set (OSFS), and the transmission $\eta_{MMF} \approx 0.92$ of the multi-mode fiber. Therefore, $\eta_t = 0.71 \times 0.56 \times 0.92 \approx 36\%$.

The total detection efficiency of the write-out channel is similar to that of the readout channel and includes the efficiency $\eta_{esp} \approx 60\%$ of light escaping from the cavity, the coupling efficiency $\eta_{SMF} \approx 71\%$ of the single-mode fiber SMF$_S$, the transmission ~56% of the optical-spectrum-filter set (OSFS), the transmission $\eta_{MMF}$ =92% of the multi-mode fiber (MMF), and the quantum efficiency $\eta_D \approx 68\%$ of the detector $D_{S_1}$ ($D_{S_2}$). Therefore, the total detection efficiency of the write-out channel is $\eta_{TD} = \eta_{esp} \times \eta_{SMF} \times \eta_{Filter} \times \eta_{MMF} \times \eta_D \approx 15\%$.

All error bars in the experimental data represent a ±1 standard deviation, which is estimated from Poissonian detection statistics using

Monte Carlo simulations.

The low total detection efficiencies can be effectively improved in future work. In this work, the cavity length reaches 6 m to obtain a long spin-wave wavelength. Such a long cavity is mainly responsible for the current cavity loss of $L \approx 13\%$. Assuming that we use optical-lattice atoms instead of the cold atoms, a short cavity can be used because the long-wavelength spin wave is no longer necessary. The cavity loss can be significantly decreased in this case. For example, it has been decreased to ~ 3.5% [21]. Assuming that the cavity loss reduces to 0.5%, the cavity escaping efficiency $\eta_{esp}$ will be up to $\sim 98\%$. Furthermore, the efficiencies $\eta_{SMF_S}$, $\eta_{Filter}$, and $\eta_{MMF}$ greatly improve to 99%, 98%, and 99%. One can also use superconductor single-photon detectors with a detection efficiency of 95% [77] instead of silicon-avalanche-photodiode single-photon detectors. With the aforementioned improvements, the total detection efficiency would change to

$$\eta_{TD} = \eta_{esp} \times \eta_{SMF_S} \times \eta_{Filter} \times \eta_{MMF} \times \eta_D \approx 0.98*0.99*0.98*0.99*0.95 \approx 90\%.$$

**The extension of the mode number.** In our presented experiment, the cavity simultaneously resonates with two spatially-distinctive write-out modes. However, the two modes are not fundamentally limited. We can greatly increase the mode number by extending the sizes of the optical elements in the cavity and aligning the added mode to propagate parallel to the original modes.

**Repeater rate for distributing an entangled photon pair over a distance $L$ through a multiplexed QR via two-photon interference.** On the basis of two-photon interference, the following expression can be used to evaluate the repeater rate for distributing an entangled photon pair over a distance $L$ through multiplexed DLCZ QR protocols with nest level $n=4$ [72]:

$$R_{rate} \approx \frac{1}{T_{cc}} P_0^{(N)} \left( \prod_{j=1}^{j=4} P_j \right) P_{pr}, \quad (2)$$

where $T_{cc} = L_0/c$ is the communication time with $c$ being the speed of light in fibers and $L_0 = L/n$ the distance between two nodes belonging to an elementary link; $P_0^{(N)} = 1-(1-P_0)^N \approx NP_0$ is the success probability for entanglement generation in an individual elementary link using multiplexed nodes that each store $N$ qubits, with $P_0 = \left( \chi^2 e^{-L_0/L_{att}} \eta_{FC}^2 \eta_{TD}^2 \right)/2$ being that using non-multiplexed nodes, $\eta_{FC}$ being the memory-to-telecom frequency conversion efficiency, $\eta_{TD}$ being the total detection efficiency, and the factor 1/2 being due to double-excitation events from a single node [31]; $P_j = \left[ \left( R_0 e^{-t_{j-1}/\tau_0} \right)^2 \eta_{TD}^2 \right]/2$ with $j=1$ to $n=4$ is the success probability for entanglement swapping at the $i$-th level, with $\tau_0$ the lifetime of the multiplexed node memory based on an atomic ensemble, and $R_0$ the qubit retrieval efficiency of the memory at zero delay; $t_0 \simeq T_{cc}/P_0^{(N)} = T_{cc}/(NP_0)$ is the time needed for the elementary entanglement generation; $t_j \simeq t_{j-1}/P_j$ $j=1$ to $n=4$ is the time needed for the $i$th-level entanglement swapping; and $P_{pr} \approx \left(R_0 e^{-t_4/\tau_0}\right)^2/2$ is the probability for distributing an entangled photon pair over the distance $L$. Using Eq. 3, we calculated the repeater rate as a function of the distance $L$ over which an entangled photon pair is distributed through a QR using nodes with CPE retrieval ($R_0$=77%) and cavity-imperfectly-enhanced (CIE) retrieval ($R_0$=58%).

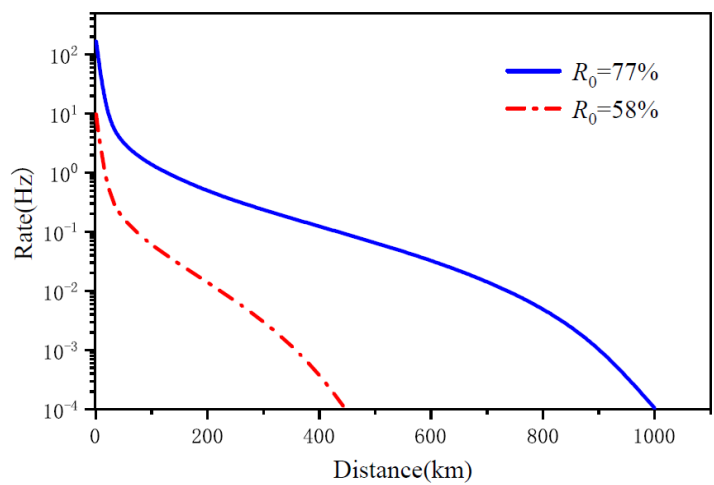

**Fig.5.** Calculated repeater rates as functions of entanglement distribution over distance $L$ according to Eq. 2, where the blue solid (red dash) curve corresponds to the quantum repeater using nodes with zero-delay retrieval efficiency $R_0$=77% ($R_0$=58%). The parameters used in the calculation are: nest level $n$=4, mode number $N$=1000, memory lifetime $\tau_0 = 16\,s$, total detection efficiency $\eta_{TD}=90\%$ for the write-out (readout) channel, and quantum frequency conversion efficiency $\eta_{FC}=33\%$.